\documentclass[12pt]{article}

\usepackage{amsmath, epsfig, cite, lineno}
\usepackage{amssymb,amsthm}
\usepackage{amsfonts, color}
\usepackage{latexsym}

\newtheorem{thm}{Theorem}

\newtheorem{conj}[thm]{Conjecture}
\newtheorem{lem}[thm]{Lemma}

\def\per{{\rm per}}

\numberwithin{equation}{section}

\begin{document}


\begin{center}
{\Large\bf Proof of a conjecture of Kl{\o}ve on permutation codes\\[5pt] under the Chebychev distance}
\end{center}

\vskip 2mm \centerline{Victor J. W. Guo$^1$ and Yiting Yang$^2$}
\begin{center}
{\footnotesize $^1$School of Mathematical Sciences, Huaiyin Normal University, Huai'an, Jiangsu 223300,
 People's Republic of China\\
{\tt jwguo@hytc.edu.cn}\\[10pt]
$^2$Department of Mathematics, Tongji University, Shanghai 200092, People's Republic of China\\
{\tt ytyang@tongji.edu.cn} }
\end{center}


\vskip 0.7cm \noindent{\bf Abstract.} Let $d$ be a positive integer and $x$ a real number.  Let $A_{d, x}$ be a $d\times 2d$ matrix
with its entries
$$
a_{i,j}=\left\{
\begin{array}{ll}
x\ \ & \mbox{for} \ 1\leqslant j\leqslant d+1-i,\\
1\ \ & \mbox{for} \ d+2-i\leqslant j\leqslant d+i,\\
0\ \ & \mbox{for} \ d+1+i\leqslant j\leqslant 2d.
\end{array}
\right.
$$
Further, let $R_d$ be a set of sequences of integers as follows:
$$R_d=\{(\rho_1, \rho_2,\ldots, \rho_d)|1\leqslant \rho_i\leqslant d+i, 1\leqslant i \leqslant d,\  \mbox{and}\  \rho_r\neq \rho_s\  \mbox{for}\  r\neq s\}.$$
and define $$\Omega_d(x)=\sum_{\rho\in R_d}a_{1,\rho_1}a_{2, \rho_2}\ldots a_{d,\rho_d}.$$

In order to give a better bound on the size of spheres of  permutation codes under the Chebychev distance,
 Kl{\o}ve introduced the above function and conjectured that $$\Omega_d(x)=\sum_{m=0}^d{d\choose m}(m+1)^d(x-1)^{d-m}.$$
 In this paper,  we settle down this conjecture positively.



\section{Introduction}

A permutation code is a subset of the symmetric group $S_n$, equipped with a distance metric. Permutation codes are of potential
use in various applications such as power-line communications and coding for flash memories used with rank modulation\cite{Jiang1,Jiang2}.
Permutation codes were extensively studied over the last decade. Hamming metric is naturally the
 first to be considered. Later,  Ulam metric\cite{Farnoud} and Kendall $\tau$-metric \cite{Buzaglo} were introduced and are now
the two most investigated metrics. However in \cite{Klove1},  a new metric named the Chebyshev metric was proposed by Kl{\o}ve et al., when they were studying the multi-level flash memory model.
A combinatorial survey on metrics related to permutations was given in \cite{Deza}.

The two main questions in coding theory are fundamental limits
on the parameters of the code (information rate versus minimum distance) and constructions
of codes that attain these limits.
It turns out that both topics are difficult for permutation codes. Few explicit constructions were obtained for various metrics and no general bounds better than the GV-bound and Sphere packing bounds were found in \cite{Barg, Buzaglo, Farnoud, Jiang1, Klove1} except for the Hamming metric\cite{Gao}. Both the GV-bound and the Sphere packing
bound  depends on the volume ($V(n, d)$) of a typical ``ball"  which consists of permutations in $S_n$
at distance at most $d$ from the identity permutation. The calculation of the volume of that ball becomes a crucial problem.

The Chebychev distance $d(p, q)$ between two permutations $p=(p_1, p_2, \ldots, p_n)$ and $q=(q_1, q_2, \ldots, q_n)$ is defined by
$$d(p, q)=\max_j|p_j-q_j|.$$

Let
$$T_{d, n}=\{p\in S_n||p_i-i|\leqslant d \ \mbox{for}\  1\leqslant i\leqslant n \}.$$ 
It is clear that $V(d, n)=|T_{d, n}|.$
The permanent of an $n\times n$ matrix $A$ is defined by
$$\per A=\sum_{p\in S_n}a_{1, p_1}\ldots a_{n, p_n}.$$
Let $A^{(d, n)}$ be the $n\times n$ matrix with $a_{i,j}^{d, n}=1$
if $|i-j|\leqslant d$ and $a_{i, j}^{d, n}=0$ otherwise. Clearly,
$V(d, n)=\per A^{(d, n)}.$ Although  the permanent
looks similar to the determinant of a matrix, it is a difficult problem
to compute the permanent for general matrices. The celebrated van
der Waerden theorem gives a lower bound for the permanent of the
so called doubly stochastic $n\times n$ matrix. Here doubly
stochastic means that all the elements are non-negative and that the
sum of the elements in any row or column is $1$. Thus, if $A$ is an
$n\times n$ matrix where the sum of the elements in any row or
column is a constant $k$, then van der Waerden's theorem gives a lower
bound on the permanent of $A$.

By noticing that most rows and columns of $A^{(d, n)}$ have the sum $2d+1$, Kl{\o}ve defined a closely related matrix $B^{(d, n)}$ with row sum and column sum $2d+1$  so that  van der Waerden's theorem can be applied. The matrix $B^{(d, n)}$ is defined as follows:

$$
b_{i,j}=\left\{
\begin{array}{ll}
0\ \ &  \ \mbox{if}\ i> j+d\ \mbox{or}\ j> i+d,\\
2\ \ &  \ \mbox{if}\ i+j\leqslant d+1\ \mbox{or}\ i+j\geqslant 2n+1-d ,\\
1\ \ &  \ \mbox{otherwise}.
\end{array}
\right.
$$
With this new defined matrix $B^{(d, n)}$, Kl{\o}ve\cite{Klove2} gave a lower bound on $V(d, n)$ as follows:
\begin{align}
V(d, n)>\frac{\sqrt{2\pi n}}{2^{2d}}\left(\frac{2d+1}{e}\right)^n. \label{eq:old}
\end{align}

Let $A_{d, 2}=(a_{i,j})$ be the upper left corner of $B^{(d, n)}$ which is a $d\times 2d$ matrix defined by
\begin{align*}
a_{i,j}=\begin{cases}
2, &\text{if $1\leqslant j\leqslant d+1-i$,}\\
1, &\text{if $d+2-i\leqslant j\leqslant d+i$,}\\
0, &\text{if $d+i+1\leqslant j\leqslant 2d$.}
\end{cases}
\end{align*}
For example,
\begin{align*}
A_{1,2}=\left(\begin{matrix}2&1\end{matrix}\right),\quad
A_{2,2}=\left(\begin{matrix}2&2&1&0\\
                            2&1&1&1\end{matrix}\right),\quad
A_{3,2}=\left(\begin{matrix}2&2&2&1&0&0\\
                            2&2&1&1&1&0\\
                            2&1&1&1&1&1   \end{matrix}\right).
\end{align*}
Let $R_d$ be a set of sequences of integers as follows:
$$R_d=\{(\rho_1, \rho_2,\ldots, \rho_d)|1\leqslant \rho_i\leqslant d+i, 1\leqslant i \leqslant d,\  \mbox{and}\  \rho_r\neq \rho_s\}.$$
Define  $$\Omega_d=\sum_{\rho\in R_d}a_{1,\rho_1}a_{2,\rho_2}\ldots a_{d,\rho_d}.$$

Let
$$\omega_d=\frac{\Omega_de^d}{(2d+1)^d}.$$
Kl{\o}ve\cite{Klove1} also gave the following lower bound on $V(n, d)$:
\begin{align}
V(d, n)>\frac{\sqrt{2\pi (n+2d)}}{\omega^2_d}\left(\frac{2d+1}{e}\right)^n. \label{eq:new}
\end{align}
Thus whether \eqref{eq:new} is an improvement compared with \eqref{eq:old} depends on the value $\Omega_d.$ Kl{\o}ve \cite{Klove2} gave the first $9$ values of $\Omega_d$ as follows:
$$3,\ 18,\ 170,\ 2200,\ 36232,\ 725200,\ 17095248,\ 463936896,\ 14246942336,$$
which coincide the sequence A074932 in \cite{Sloane},
and made the following conjecture.

\begin{conj}\label{constant}{\rm\cite[Conjecture 1]{Klove2}}For any positive integer $d$,
$$\Omega_d=\sum_{m=0}^d{d\choose m}(m+1)^d.$$
\end{conj}

Kl{\o}ve showed that the equation \eqref{eq:new} improves on \eqref{eq:old} if Conjecture \ref{constant} is true.
Furthermore, let $A_{d,x}=(a_{i,j})$ be the $d\times 2d$ matrix defined by
\begin{align*}
a_{i,j}=\begin{cases}
x, &\text{if $1\leqslant j\leqslant d+1-i$,}\\
1, &\text{if $d+2-i\leqslant j\leqslant d+i$,}\\
0, &\text{if $d+i+1\leqslant j\leqslant 2d$.}
\end{cases}
\end{align*}
 and  let
$$\Omega_d(x)=\sum_{\rho\in R_d}a_{1, \rho_1}a_{2, \rho_2}\ldots a_{d,\rho_d}.$$
In particular, $\Omega_d(2)=\Omega_d.$  Kl{\o}ve
gave the following generalized conjecture and verified it
for $d\leqslant 9.$

\begin{conj}\label{variable}{\rm\cite[Conjecture 2]{Klove2}}For any positive integer $d$,
\begin{align}
\Omega_d(x)=\sum_{m=0}^d{d\choose m}(m+1)^d(x-1)^{d-m}.\label{eq:omega}
\end{align}
\end{conj}

In this paper, we shall prove that Conjecture \ref{variable} is true.

\section{Proof of Kl{\o}ve's Conjecture}

\begin{thm}\label{thm:main}
For any positive integer $d$, the identity \eqref{eq:omega} holds.
\end{thm}

Actually, for any $m\times n$ matrix $A=(a_{i,j})$ with $m\leqslant n$, the permanent function of $A$ is already defined as follows (see, for example, \cite{Ryser}):
\begin{align*}
\per(A)=\sum_{\sigma\in P(n,m)}a_{1,\sigma_1}a_{2,\sigma_2}\cdots a_{m,\sigma_m},
\end{align*}
where $P(n,m)$ denotes the set of all $m$-permutations of the $n$-set $\{1,2,\ldots,n\}$.


In fact, by the definition of $R_d$, we know that $R_d$ is exactly the subset of all $d$-permutations of the $2d$-set $\{1,2,\ldots,2d\}$ such that $\sigma\in R_d$ if and only if $a_{1,\sigma_1}a_{2,\sigma_2}\cdots a_{d,\sigma_d}\neq 0.$
Hence we have $\Omega_d(x)=\per(A_{d,x})$.

In order to prove Theorem \ref{thm:main}, we first give a related combinatorial identity.
\begin{lem}\label{lem:main}
Let $m$ and $n$ be positive integers. Then
\begin{align}
\sum_{1\leqslant k_1\leqslant k_2\leqslant\cdots\leqslant k_m\leqslant n}
\prod_{i=0}^{m} k_{i}(n+m-i)^{k_{i+1}-k_{i}}={n+m-1\choose m}n^{n+m-1}, \label{eq:lem}
\end{align}
where $k_0=1$ and $k_{m+1}=n$.
\end{lem}
For example, we have
\begin{align*}
\sum_{1\leqslant k_1\leqslant k_2\leqslant k_3\leqslant n}k_1 k_2 k_3(n+3)^{k_1-1}(n+2)^{k_2-k_1}(n+1)^{k_3-k_2}n^{n-k_3}
={n+2\choose 3}n^{n+2}.
\end{align*}
\noindent{\it Proof of Lemma {\rm\ref{lem:main}.}} We compute the multiple sum in the order from $k_m$ to $k_1$. It can be proved by induction on $k_{m-1},k_{m-2},
\ldots,k_{m-i-1}$ respectively that
\begin{align}
&\sum_{k_m=k_{m-1}}^n k_m (n+1)^{k_{m}-k_{m-1}}n^{n-k_{m}}=(n-k_{m-1}+1)n^{n-k_{m-1}+1}, \notag\\
&\sum_{k_{m-1}=k_{m-2}}^n k_{m-1}(n-k_{m-1}+1) (n+2)^{k_{m-1}-k_{m-2}}n^{n-k_{m-1}+1}={n-k_{m-2}+2\choose 2}n^{n-k_{m-2}+2}, \notag\\
&\cdots, \notag\\
&\sum_{k_{m-i}=k_{m-i-1}}^n k_{m-i}{n-k_{m-i}+i\choose i}(n+i+1)^{k_{m-i}-k_{m-i-1}}n^{n-k_{m-i}+i} \notag\\
&\quad{}={n-k_{m-i-1}+i+1\choose i+1}n^{n-k_{m-i-1}+i+1}.  \label{eq:multi}
\end{align}
By choosing $i=m-1$ in \eqref{eq:multi}, we complete the proof of \eqref{eq:lem}. \qed

\medskip
\noindent{\it Proof of Theorem {\rm\ref{thm:main}.}} It is clear that \eqref{eq:omega} is equivalent to
\begin{align}
\Omega_d(x+1)=\sum_{m=0}^d{d\choose m}(d-m+1)^d x^m . \label{eq:omega-2}
\end{align}
Therefore, it suffices to show that the coefficient $b_m$ of $x^m$ in $\Omega_d(x+1)$ is equal to
${d\choose m}(d-m+1)^d$. By the definition of $\Omega_d(x+1)$, we know that each $x$ comes from the first term in $x+1$.

To compute $b_m$, we first choose $m$ $x$'s from $m$ $(x+1)$'s which are not in the same row nor in the same column of
the matrix $A_{d,x+1}$, and then choose $(d-m)$ $1$'s in the other $d-m$ rows so that no $1$'s are in the same column.
Suppose that the $m$ $x$'s are chosen from the rows indexed by $d+1-i_1,d+1-i_2,\ldots,d+1-i_m$ with $i_1<i_2<\cdots<i_m$, respectively.
By noticing that the $(d+1-i)$-th row has $i$ $(x+1)$'s and all the $x$'s we choose must be in different columns, we have
$i_1(i_2-1)(i_3-2)\cdots (i_m-m+1)$ ways to do this. As for the number of ways to choose 1's in the remaining rows, we notice that the $i$-th row has $d+i$ 1's including those $1$'s in $(x+1)$'s and all these $1$'s form several right trapezoids in the matrix $A_{d,x+1}$. Therefore,
there are $(d+1)^{i_1-1}d^{i_2-i_1-1}(d-1)^{i_3-i_2-1}\cdots (d-m+1)^{d-i_m}$ ways to choose the remaining $1$'s.
It follows that
\begin{align*}
b_m&=\sum_{1\leqslant i_1<i_2<\cdots <i_m\leqslant d}i_1(i_2-1)(i_3-2)\cdots (i_m-m+1)\\
&\qquad\qquad{}\times(d+1)^{i_1-1}d^{i_2-i_1-1}(d-1)^{i_3-i_2-1}\cdots (d-m+1)^{d-i_m} \\
&=\sum_{1\leqslant k_1\leqslant k_2\leqslant\cdots\leqslant k_m\leqslant d-m+1}
\prod_{i=0}^{m} k_{i}(d+1-i)^{k_{i+1}-k_{i}},
\end{align*}
where $k_s=i_s-s+1$ ($s=1,\ldots,m$), $k_0=1$, and $k_{m+1}=d-m+1$. By replacing $n$ by $d-m+1$ in \eqref{eq:lem}, we obtain
$b_m={d\choose m}(d-m+1)^d$. This completes the proof.  \qed

\vskip 5mm \noindent{\bf Acknowledgments.} The authors thank the anonymous referees for their helpful comments on a previous version of this paper.
The first author was partially supported by the National Natural
Science Foundation of China under Grant No.~11371144 and the Qing Lan Project of Jiangsu Province. The second author was partially  supported by
the National Natural Science Foundation of China under Grant No.~11101360 and Outstanding Young Scholar Foundation
of Tongji University under Grant No. 2013KJ031.

\end{document}